\documentclass[sigconf]{acmart}

\AtBeginDocument{%
  \providecommand\BibTeX{{%
    \normalfont B\kern-0.5em{\scshape i\kern-0.25em b}\kern-0.8em\TeX}}}

\setcopyright{acmcopyright}
\copyrightyear{2025}
\acmYear{2025}
\setcopyright{rightsretained}
\acmConference[UIST Adjunct '25]{The 38th Annual ACM Symposium on User Interface Software and Technology}{September 28--October 1, 2025}{Busan, Republic of Korea}
\acmBooktitle{The 38th Annual ACM Symposium on User Interface Software and Technology (UIST Adjunct '25), September 28--October 1, 2025, Busan, Republic of Korea}
\acmISBN{979-8-4007-2036-9/25/09}
\acmDOI{10.1145/3746058.3758451}

\usepackage{amsmath}
\usepackage{physics}
\usepackage{array}

\usepackage{enumitem}
\usepackage{hyperref}
\usepackage{hyperxmp}

\usepackage{xcolor}
\usepackage{caption}
\definecolor{myblue}{rgb}{0.216,0.494,0.965}

\begin{document}

\title[GhostObjects]{GhostObjects: Instructing Robots by Manipulating Spatially Aligned Virtual Twins in Augmented Reality}

\author{Lauren W. Wang}
\affiliation{
  \institution{Princeton University}
  \streetaddress{TBD}
  \city{Princeton, New Jersey}
  \country{USA}}
\email{wlauren@princeton.edu}

\author{Parastoo Abtahi}
\affiliation{
  \institution{Princeton University}
  \streetaddress{TBD}
  \city{Princeton, New Jersey}
  \country{USA}}
\email{parastoo@princeton.edu}

\renewcommand{\shortauthors}{Wang et al.}

\begin{abstract}
    Robots are increasingly capable of autonomous operations, yet human interaction remains essential for issuing personalized instructions. Instead of directly controlling robots through Programming by Demonstration (PbD) or teleoperation, we propose giving instructions by interacting with \textit{GhostObjects}—world-aligned, life-size virtual twins of physical objects—in augmented reality (AR). By direct manipulation of \textit{GhostObjects}, users can precisely specify physical goals and spatial parameters, with features including real-world lasso selection of multiple objects and snapping back to default positions, enabling tasks beyond simple pick-and-place. 
\end{abstract}



\maketitle

\section{Introduction}
    We are increasingly surrounded by autonomous robots that augment our capabilities in a variety of applications, including household cleaning, package delivery, and home furnishing. These contexts require users to give explicit instructions on specific objects at a precise location in space (e.g., where to put which clothes, which shelf to load the unboxed products, where to hang the painting).

    Augmented reality (AR) has emerged as a promising approach for human-robot interaction that requires spatial manipulation, including end-user robot programming that modifies waypoints or trajectories for robot motion planning~\cite{Gadre2019, Quintero2018, Fang2012, Lunding2024, ikeda2025marcer, cao2019v}, improving teleoperation of robotic arms and drones through AR visualizations~\cite{Hedayati2018, chen20243d}, and communicating robot intent~\cite{rosen2020communicating, zhu2017augmented, Walker2018}. 

    \begin{figure}[t!]
        \includegraphics[width=\linewidth]{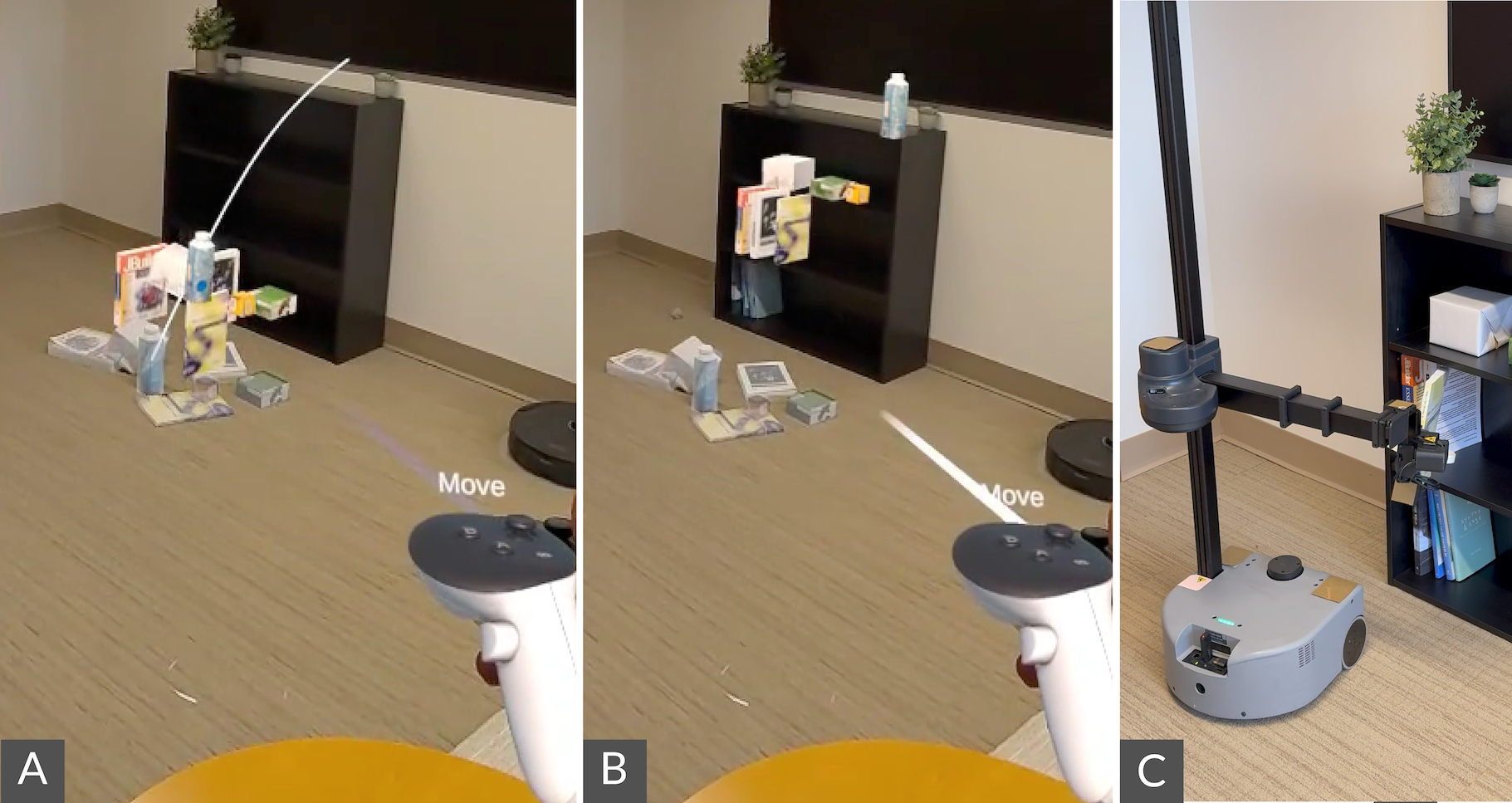}
        \vspace{-14pt}
        \caption{Moving a pile of \textit{GhostObjects} by (A) directly manipulating the world-aligned, life-size virtual twins of the real objects and (B) snapping them onto the shelf by releasing along the default trajectory. (C) A robot receives the instruction and carries out the task.}
        \label{fig:teaser}
        \Description{(A) The user selects a pile of world-aligned, life-size virtual twins of the real objects on the floor next to a shelf. (B) The user moves the controller while holding the trigger to move the GhostObjects. The user snaps them onto the shelf by releasing the controller trigger along the default trajectory. (C) A robot receives the instruction and carries out the task by putting a real book back on the shelf.}
        \vspace{-5pt}
    \end{figure}

    Prior work has explored a wide range of methods for instructing robots. GhostAR~\cite{Cao2019} uses demonstrative role-playing, an embodied authoring approach that requires a high user understanding of robot capabilities and demonstration costs. Rather than relying on captured human motion, SayCan~\cite{saycan2022arxiv} employs a textual conversational agent equipped with high-level semantic knowledge to perform generalized tasks. Yet natural language inputs are known to suffer from issues of discoverability and ambiguity. Marcer~\cite{ikeda2025marcer} also uses language commands, but introduces AR to visualize virtual twins of robots and objects for clarity. Still, these AR visualizations aren't interactive, so the issues with language inputs remain. ImageInThat~\cite{mahadevan2025imageinthat} reduces this friction by enabling direct manipulation of pixels though dragging corresponding objects in an image editor. However, it remains limited to a 2D interface. HoloSpot~\cite{pablo2024} moves into the 3D world, enabling users to drag and drop a single virtual object to issue pick-and-place commands to the robot in a miniaturized world that is not co-located. Its World-in-Miniature (WiM) setup, while effective for remote teleoperation, is not ideal when robot instructions demand co-located viewing, as in scenarios like arranging items in new locations or deciding where to hang a painting for the best view. X-HR Training~\cite{Wang2023XHRTraining} similarly showed examples of human-robot collaboration in 3D AR, but direct manipulation of virtual twins was used for Learning by Demonstration purposes without physical counterparts, and was not spatially aligned with real objects. In other examples, co-located virtual objects served as non-interactive AR visualizations for previewing instruction outcomes.\looseness=-1
    
    Instead of directly controlling robots through expert authoring, our approach centers on interacting with target objects. Rather than relying on language commands, we enable direct manipulation for greater precision. To support co-located tasks and life-size manipulation, we develop a 3D, aligned AR environment. In sum, we propose instructing robots by directly interacting with virtual twins of real-world objects--\textit{GhostObjects}--in spatially aligned physical space using augmented reality (AR). Users can raycast to select an object in passthrough view and directly manipulate the \textit{GhostObject}. In Figure~\ref{fig:teaser}, to instruct a robot to move a pile of items, the user can grab the \textit{GhostObjects} and place them freely within a physics-enabled environment or snap them back to default locations.\looseness=-1

\section{Ghost Objects}


\subsection{Single-object Raycast Selection}
    Unlike traditional interfaces (e.g. WiM~\cite{pablo2024}) that require users to mentally map targets into digital space, our system enables direct AR interaction with co-located objects. To select an object, the user initiates a raycast into the environment by pointing the controller. Pressing the trigger button selects the first object intersected by the ray.\looseness=-1

\subsection{Multi-object Lasso Selection}
\begin{figure}[h!]
        \includegraphics[width=\linewidth]{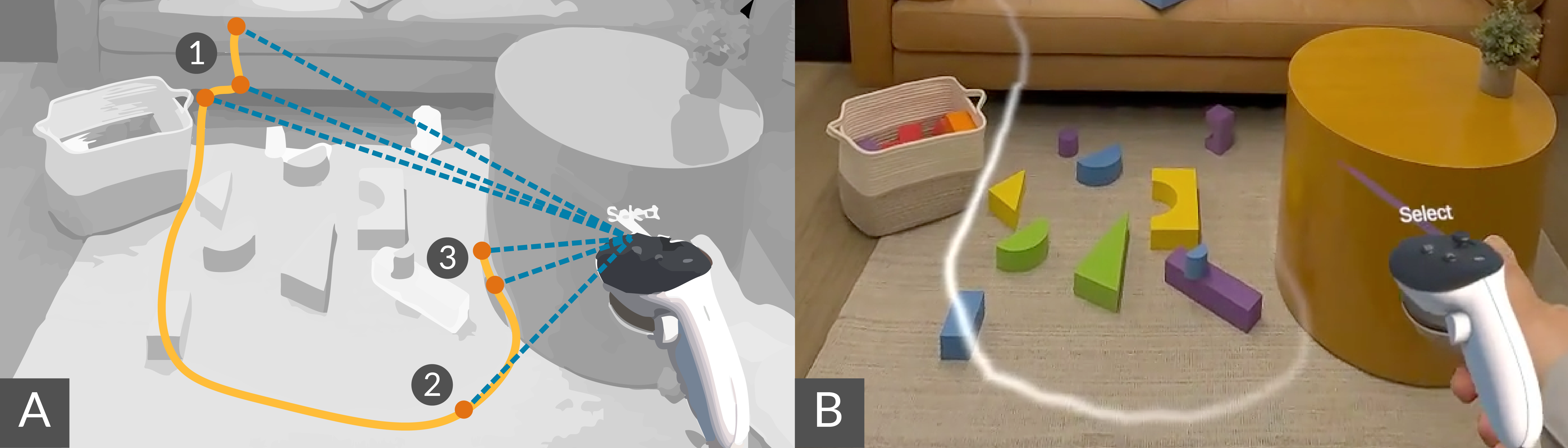}
        \vspace{-16pt}
        \caption{Lasso selection. (A) An illustration showing the lasso intersecting the sofa (1), carpet (2), and table (3). The yellow line represents the lasso, and the blue line represents the raycast. (B) The AR headset view of the lasso selection.}
        \label{fig:lasso}
        \Description{(A) The user draws a lasso around the foam blocks scattered on the floor using a ray from the controller. A gray-scaled image with an annotated lasso trajectory highlights the path the user lassoed with the controller. (B) The image shows the same lasso selection setup but in the real AR headset view.}
    \end{figure}

   Additionally, users can select multiple objects by holding the trigger button and drawing a lasso to outline an arbitrary shape around their desired targets. The ray projected from the controller intersects with surfaces in the environment (Figure~\ref{fig:lasso}A). As the user moves the controller, intersection points connect, forming a closed boundary when the trigger is released. This lassoed boundary creates a volume extending towards the controller, and objects that overlap this volume are grouped and selected together.\looseness=-1

\subsection{Spatially Aligned Virtual Twins}
    When an object is selected, a virtual twin—called a \textit{GhostObject}—is immediately overlaid on the real object in AR. \textit{GhostObjects} are life-sized and spatially aligned with the real world. We utilize Meta Quest’s Space Setup feature, which scans the scene and generates bounding boxes for spatial anchors. We then leverage Meta’s Mixed Reality Utility Kit (MRUK) to access the positions of these spatial anchors, which serve as reference points for spawning \textit{GhostObjects} in their relative poses. This ensures that objects consistently appear in the correct physical location, eliminating repeated manual calibration. The virtual twins provide users with immediate feedback and help them visualize the anticipated outcomes of robot tasks.

\subsection{Direct Manipulation of \textit{GhostObjects}}
    \textit{GhostObjects} can be directly manipulated, minimizing the gulf between users’ intentions and task execution~\cite{Norman1985}. Direct manipulation lets \textit{GhostObjects} be moved, scaled, aligned, and visually modified in AR. Users can hold the trigger button to grab, move, and place \textit{GhostObjects}, which conform to the environment’s simulated physics. Beyond relocation, users can also modify \textit{GhostObjects} that change form or state—for example, adjusting the liquid level inside a bottle during filling tasks or compressing empty boxes for recycling.\looseness=-1

\subsection{Snap to Default}
    Each \textit{GhostObject} has a customizable default location that stores the pose where the object is expected to be. When moved, an arched trajectory line appears from its current position to the default location. Releasing the \textit{GhostObject} along this path causes it to snap back to default. Users can redefine the default location by placing the \textit{GhostObject} elsewhere and saving that as the default.

\section{Application Scenarios}
\subsection{Home Tidying}
    \begin{figure}[h!]
        \includegraphics[width=\linewidth]{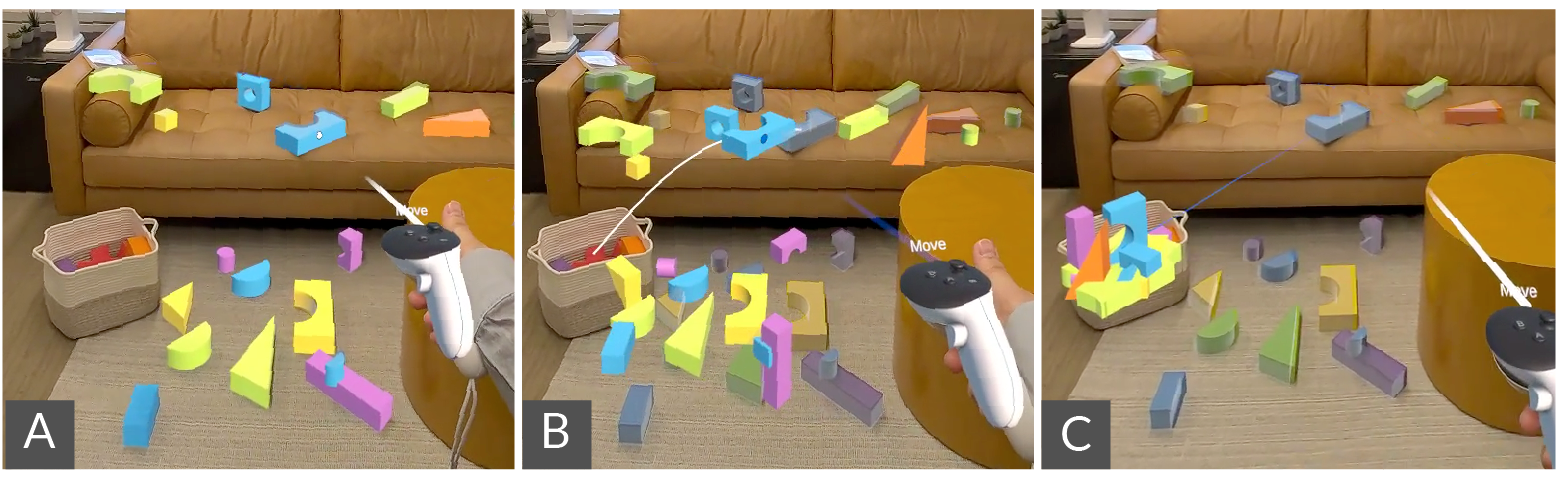}
        \vspace{-16pt}
        \caption{Relocating scattered foam blocks to a basket using direct manipulation and the snap-to-default feature.}
        \label{fig:app_foam}
        \Description{(A) The user points at foam blocks scattered on the floor and couch using a ray from the controller. (B) The selected blocks turn into GhostObjects and begin snapping to their default positions inside a basket. (C) The user uses the controller to directly manipulate the group of GhostObjects toward the basket. All blocks maintain their color-coded appearance during manipulation.}
    \end{figure}
    We demonstrate a walkthrough of a domestic scenario where a user instructs a robot to move foam blocks into a basket beside the sofa. After lasso selecting the blocks (Figure~\ref{fig:lasso}B), she points the controller at one of the \textit{block GhostObjects} and holds the trigger to grab it (Figure~\ref{fig:app_foam}A). Since all the blocks are selected, they move as a group, with a trajectory line connecting them to the basket to indicate their default location (Figure~\ref{fig:app_foam}B). She moves the blocks along this path and releases the trigger, snapping them into the basket (Figure~\ref{fig:app_foam}C).\looseness=-1

\subsection{Filling Bottle}
    \begin{figure}[h!]
        \includegraphics[width=\linewidth]{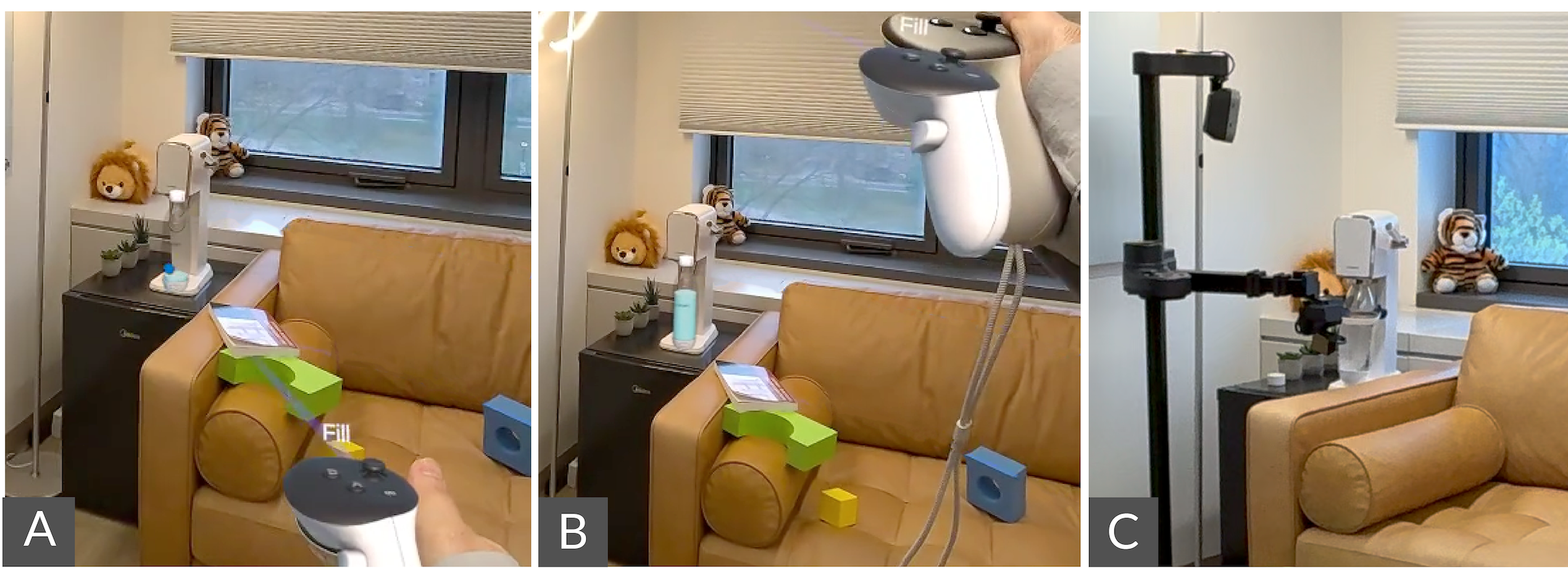}
        \vspace{-16pt}
        \caption{Filling \textit{water GhostObject} to a desired level.}
        \label{fig:app_fillsoda}
        \Description{(A) The user selects the "Fill" action from a radial menu targeting a soda bottle. (B) The user raises the controller upward while holding the trigger. The liquid level in the virtual bottle increases, showing the effect of direct manipulation for representing spatially encoded object states. (C) A robot picks up the filled bottle.}
    \end{figure}
    
    While we focus on pick-and-place actions, \textit{GhostObjects} can also extend to robot tasks involving state change or deformation. For example, the user finds a soda machine beside the sofa and wants the robot to fill the bottle with water. She holds the trigger and raises her hand up, deforming the \textit{water GhostObject} to specify the desired fill level (Figure~\ref{fig:app_fillsoda}). Once the user’s instructions are complete, the action and spatial parameters are sent to the robot.

\section{Conclusion and Future Work}
    In this work, we present \textit{GhostObjects}—world-aligned, life-size virtual twins of physical objects in augmented reality (AR)—that can be directly manipulated as a means of instructing robots.
    In the future, we plan to explore various object affordances and robot tasks to further experiment with \textit{GhostObjects}. We also aim to leverage modern reconstruction techniques and vision-language models to automate the creation of \textit{GhostObjects}, enhancing their generalizability. Additionally, we intend to conduct further evaluations of the efficacy of using \textit{GhostObjects} in real-world robot tasks.

\bibliographystyle{ACM-Reference-Format}
\bibliography{References}

\end{document}